\documentclass[11pt]{article}
\usepackage[a4paper]{geometry}
\usepackage{amsfonts, amsmath, amssymb, amsthm, graphicx, caption, authblk, multirow, makecell, framed, float, xcolor, enumitem, tikz, hyperref}
\theoremstyle{definition}

\theoremstyle{remark}

\newcommand*{\mybox}[1]{%
  \framebox{\raisebox{0cm}[0.5\baselineskip][0.05\baselineskip]{%
    \hbox to 0.1cm {\hss#1\hss}}}\hspace{0.05cm}}

\begin{document}
\title{Using Five Cards to Encode Each Integer in $\mathbb{Z}/6\mathbb{Z}$}
\author[1]{Suthee Ruangwises\thanks{\texttt{ruangwises@gmail.com}}}
\affil[1]{Department of Mathematical and Computing Science, Tokyo Institute of Technology, Tokyo, Japan}
\date{}
\maketitle

\begin{abstract}
Research in secure multi-party computation using a deck of playing cards, often called card-based cryptography, dates back to 1989 when Den Boer introduced the ``five-card trick'' to compute the logical AND function. Since then, many protocols to compute different functions have been developed. In this paper, we propose a new encoding scheme that uses five cards to encode each integer in $\mathbb{Z}/6\mathbb{Z}$. Using this encoding scheme, we develop protocols that can copy a commitment with 13 cards, add two integers with 10 cards, and multiply two integers with 14 cards. All of our protocols are the currently best known protocols in terms of the required number of cards. Our encoding scheme can be generalized to encode integers in $\mathbb{Z}/n\mathbb{Z}$ for other values of $n$ as well.

\end{abstract}

\section{Introduction}
Secure multi-party computation, one of the most actively studied areas in cryptography, involves situations where multiple parties want to compare their secret information without revealing it. Many researchers focus on developing secure multi-party computation protocols using physical objects such as a deck of playing cards, creating a research area often called \textit{card-based cryptography}. The benefit of card-based protocols is that they provide simple solutions to real-world situations using only objects found in everyday life without requiring computers. Moreover, these intuitive protocols are easy to understand and verify the correctness and security, even for non-experts in cryptography, and thus can be used for educational purposes to teach the concept of secure multi-party computation.

\subsection{Protocols of Boolean Functions}
Research in card-based cryptography dates back to 1989 when Den Boer \cite{denboer} proposed a protocol called the \textit{five-card trick} to compute the logical AND function on two players' bits $a$ and $b$.

The five-card trick protocol uses three identical \mybox{$\clubsuit$}s and two identical \mybox{$\heartsuit$}s, with all cards having indistinguishable back sides. We use a \textit{commitment} \mbox{\mybox{$\clubsuit$}\mybox{$\heartsuit$}} to encode 0, and a commitment \mbox{\mybox{$\heartsuit$}\mybox{$\clubsuit$}} to encode 1. First, each player is given one \mybox{$\clubsuit$} and one \mybox{$\heartsuit$}, and another \mybox{$\clubsuit$} is put face-down on a table. The first player places his/her commitment of $a$ face-down to the left of the \mybox{$\clubsuit$} on the table, while the second player places his/her commitment of $b$ face-down to the right of it. Then, we swap the two cards in the commitment of $b$, resulting in the following four possible sequences.

\begin{figure}[H]
    \centering
		\begin{minipage}{3cm}
				\centering
				$(a,b)=(0,0):$\\~\\
				$(a,b)=(0,1):$\\~\\
				$(a,b)=(1,0):$\\~\\
				$(a,b)=(1,1):$
		\end{minipage}
    \begin{minipage}{3cm}
        \centering
        \mybox{$\clubsuit$} \mybox{$\heartsuit$} \mybox{$\clubsuit$} \mybox{$\clubsuit$} \mybox{$\heartsuit$}\\~\\
				\mybox{$\clubsuit$} \mybox{$\heartsuit$} \mybox{$\clubsuit$} \mybox{$\heartsuit$} \mybox{$\clubsuit$}\\~\\
				\mybox{$\heartsuit$} \mybox{$\clubsuit$} \mybox{$\clubsuit$} \mybox{$\clubsuit$} \mybox{$\heartsuit$}\\~\\
				\mybox{$\heartsuit$} \mybox{$\clubsuit$} \mybox{$\clubsuit$} \mybox{$\heartsuit$} \mybox{$\clubsuit$}
		\end{minipage}
		\begin{minipage}{0.3cm}
				\centering
				$\Rightarrow$\\~\\
				$\Rightarrow$\\~\\
				$\Rightarrow$\\~\\
				$\Rightarrow$
		\end{minipage}
		\begin{minipage}{3cm}
        \centering
        \mybox{$\clubsuit$} \mybox{$\heartsuit$} \mybox{$\clubsuit$} \mybox{$\heartsuit$} \mybox{$\clubsuit$}\\~\\
				\mybox{$\clubsuit$} \mybox{$\heartsuit$} \mybox{$\clubsuit$} \mybox{$\clubsuit$} \mybox{$\heartsuit$}\\~\\
				\mybox{$\heartsuit$} \mybox{$\clubsuit$} \mybox{$\clubsuit$} \mybox{$\heartsuit$} \mybox{$\clubsuit$}\\~\\
				\mybox{$\heartsuit$} \mybox{$\clubsuit$} \mybox{$\clubsuit$} \mybox{$\clubsuit$} \mybox{$\heartsuit$}
		\end{minipage}
\end{figure}

Among all cases, there are only two possible sequences in a cyclic rotation of the deck: \mbox{\mybox{$\heartsuit$}\mybox{$\clubsuit$}\mybox{$\heartsuit$}\mybox{$\clubsuit$}\mybox{$\clubsuit$}} and \mbox{\mybox{$\heartsuit$}\mybox{$\heartsuit$}\mybox{$\clubsuit$}\mybox{$\clubsuit$}\mybox{$\clubsuit$}}, with the latter occurring if and only if $a=b=1$. We can hide the initial position of the cards by applying a \textit{random cut} to shift the sequence into a uniformly random cyclic shift, i.e. a permutation uniformly chosen at random from $\{\text{id}, \pi, \pi^2, \pi^3, \pi^4\}$ where $\pi = \text{(1 2 3 4 5)}$, before turning all cards face-up. Hence, we can determine whether $a \wedge b = 1$ without leaking any other information.

Since the introduction of the five-card trick, several other protocols to compute the AND function have been developed. These subsequent protocols \cite{abe,abe2,crepeau,isuzugawa,koch2,koch,mizuki16,mizuki12,mizuki09,niemi,ruangwises,stiglic} either reduced the number of required cards or improved properties of the protocol involving output format, type of shuffles, running time, etc.

Apart from AND protocols, protocols to compute other Boolean functions have also been developed, such as logical XOR protocols \cite{crepeau,mizuki09,mizuki06,toyoda2}, copy protocols \cite{crepeau,koyama,mizuki09} (duplicating a commitment), \textit{majority function} protocols \cite{nishida2,toyoda} (deciding whether there are more 1s than 0s in the inputs), \textit{equality function} protocols \cite{ruangwises2,shinagawa} (deciding whether all inputs are equal), and a voting protocol \cite{mizuki13} (adding bits and storing the sum in binary representation).

Nishida et al. \cite{nishida} proved that any $n$-variable Boolean function can be computed with $2n+6$ cards, and any such function that is symmetric can be computed with $2n+2$ cards.

\subsection{Protocols of Non-Boolean Functions}
While almost all of the existing protocols were designed to compute Boolean functions, a few results also focused on computing functions in $\mathbb{Z}/n\mathbb{Z}$ for $n>2$. Shinagawa et al. \cite{polygon} used a regular $n$-gon card to encode each integer in $\mathbb{Z}/n\mathbb{Z}$ and proposed a copy protocol and an addition protocol for integers in $\mathbb{Z}/n\mathbb{Z}$. Their encoding scheme can be straightforwardly converted to the one using regular cards. In another result, Shinagawa and Mizuki \cite{triangle} developed a protocol to multiply two integers in $\mathbb{Z}/3\mathbb{Z}$ using triangle cards. Their idea can also be generalized to multiply integers in $\mathbb{Z}/n\mathbb{Z}$ using regular cards.

Another straightforward method to compute functions on $\mathbb{Z}/n\mathbb{Z}$ is to convert each integer in $\mathbb{Z}/n\mathbb{Z}$ into its binary representation and encode each digit with two cards, resulting in the total of $2\lceil \lg n \rceil$ cards, and then apply the protocol of Nishida et al. \cite{nishida} to compute these functions.

\subsection{Our Contribution}
In this paper, we propose a new encoding scheme that uses five cards to encode each integer in $\mathbb{Z}/6\mathbb{Z}$. The idea behind this scheme is to use the first two cards and the last three cards to represent its residues in modulo 2 and modulo 3, respectively, and then use the converted scheme of Shinagawa et al. \cite{polygon} to encode each part. This simple trick significantly reduces the number of required cards for every basic protocol, which is the main objective of developing card-based protocols. Using this encoding scheme, we present protocols that can copy a commitment with 13 cards, add two integers with ten cards, and multiply two integers with 14 cards. These three protocols are the essential ones that enable us to compute any polynomial function $f: (\mathbb{Z}/6\mathbb{Z})^k \rightarrow \mathbb{Z}/6\mathbb{Z}$. All of these three protocols are the currently best known ones in terms of the required number of cards (see Table \ref{table1}).

Our encoding scheme can be generalized to other rings of integers modulo $n$, including $\mathbb{Z}/12\mathbb{Z}$ where our protocols are the currently best known ones as well.

\begin{table}
	\centering
	\begin{tabular}{|c|c|c|c|}
		\hline
		\multirow{2}{*}{\textbf{Encoding Scheme}} & \multicolumn{3}{c|}{\textbf{Number of required cards}} \\ \cline{2-4}
		& \textbf{Copy} & \textbf{Addition} & \textbf{Multiplication} \\ \hline
		\textbf{Shinagawa et al. \cite{polygon}} & 18 & 12 & 42 \\ \hline
		\textbf{Nishida et al. \cite{nishida}} & 14 & 22 & 22 \\ \hline
		\textbf{Our scheme (\S 5)} & \textbf{13} & \textbf{10} & \textbf{14} \\ \hline
	\end{tabular}
	\medskip
	\caption{The number of required cards for copy, addition, and multiplication protocols in $\mathbb{Z}/6\mathbb{Z}$ using each encoding scheme} \label{table1}
\end{table}

\section{Preliminaries} \label{prelim}
\subsection{Sequence of Cards}
For $0 \leq a < n$, define $E_n(a)$ to be a sequence of consecutive $n$ cards, with all of them being \mybox{$\heartsuit$} except the $(a+1)$-th card from the left being \mybox{$\clubsuit$}, e.g. $E_4(2)$ is \mbox{\mybox{$\heartsuit$}\mybox{$\heartsuit$}\mybox{$\clubsuit$}\mybox{$\heartsuit$}}. Unless stated otherwise, the cards in $E_n(a)$ are arranged horizontally as defined above. In some situations, however, we may arrange the cards vertically, with the leftmost card becoming the topmost card and the rightmost card becoming the bottommost card.

Many existing protocols use the sequence $E_n(a)$ to encode an integer $a$ in $\mathbb{Z}/n\mathbb{Z}$, such as millionaire protocols \cite{yao,yao2}, a ranking protocol \cite{takashima}, and protocols of zero-knowledge proof for logic puzzles \cite{makaro,suguru,numberlink,ripple,bridges}.

\subsection{Matrix}
In an $m \times n$ matrix of cards, let Row $i$ ($0 \leq i < m$) denote the $(i+1)$-th topmost row of the matrix, and Column $j$ ($0 \leq j < n$) denote the $(j+1)$-th leftmost column of the matrix.

\subsection{Pile-Shifting Shuffle} \label{shift}
\begin{figure}[H]
\begin{center}
\begin{tikzpicture}
\node at (0.00,0.60) {\mybox{?}};
\node at (0.50,0.60) {\mybox{?}};
\node at (1.00,0.60) {\mybox{?}};
\node at (1.50,0.60) {\mybox{?}};
\node at (2.00,0.60) {\mybox{?}};
\node at (2.50,0.60) {\mybox{?}};

\node at (0.00,1.20) {\mybox{?}};
\node at (0.50,1.20) {\mybox{?}};
\node at (1.00,1.20) {\mybox{?}};
\node at (1.50,1.20) {\mybox{?}};
\node at (2.00,1.20) {\mybox{?}};
\node at (2.50,1.20) {\mybox{?}};

\node at (0.00,1.80) {\mybox{?}};
\node at (0.50,1.80) {\mybox{?}};
\node at (1.00,1.80) {\mybox{?}};
\node at (1.50,1.80) {\mybox{?}};
\node at (2.00,1.80) {\mybox{?}};
\node at (2.50,1.80) {\mybox{?}};

\node at (0.00,2.40) {\mybox{?}};
\node at (0.50,2.40) {\mybox{?}};
\node at (1.00,2.40) {\mybox{?}};
\node at (1.50,2.40) {\mybox{?}};
\node at (2.00,2.40) {\mybox{?}};
\node at (2.50,2.40) {\mybox{?}};

\node at (-0.40,0.60) {3};
\node at (-0.40,1.20) {2};
\node at (-0.40,1.80) {1};
\node at (-0.40,2.40) {0};

\node at (0.00,2.90) {0};
\node at (0.50,2.90) {1};
\node at (1.00,2.90) {2};
\node at (1.50,2.90) {3};
\node at (2.00,2.90) {4};
\node at (2.50,2.90) {5};

\node at (3.40,1.50) {\LARGE{$\Rightarrow$}};
\end{tikzpicture}
\begin{tikzpicture}
\node at (0.00,0.60) {\mybox{?}};
\node at (0.50,0.60) {\mybox{?}};
\node at (1.00,0.60) {\mybox{?}};
\node at (1.50,0.60) {\mybox{?}};
\node at (2.00,0.60) {\mybox{?}};
\node at (2.50,0.60) {\mybox{?}};

\node at (0.00,1.20) {\mybox{?}};
\node at (0.50,1.20) {\mybox{?}};
\node at (1.00,1.20) {\mybox{?}};
\node at (1.50,1.20) {\mybox{?}};
\node at (2.00,1.20) {\mybox{?}};
\node at (2.50,1.20) {\mybox{?}};

\node at (0.00,1.80) {\mybox{?}};
\node at (0.50,1.80) {\mybox{?}};
\node at (1.00,1.80) {\mybox{?}};
\node at (1.50,1.80) {\mybox{?}};
\node at (2.00,1.80) {\mybox{?}};
\node at (2.50,1.80) {\mybox{?}};

\node at (0.00,2.40) {\mybox{?}};
\node at (0.50,2.40) {\mybox{?}};
\node at (1.00,2.40) {\mybox{?}};
\node at (1.50,2.40) {\mybox{?}};
\node at (2.00,2.40) {\mybox{?}};
\node at (2.50,2.40) {\mybox{?}};

\node at (-0.40,0.60) {3};
\node at (-0.40,1.20) {2};
\node at (-0.40,1.80) {1};
\node at (-0.40,2.40) {0};

\node at (0.00,2.90) {2};
\node at (0.50,2.90) {3};
\node at (1.00,2.90) {4};
\node at (1.50,2.90) {5};
\node at (2.00,2.90) {0};
\node at (2.50,2.90) {1};
\end{tikzpicture}
\caption{An example of a pile-shifting shuffle on a $4 \times 6$ matrix}
\label{fig3}
\end{center}
\end{figure}
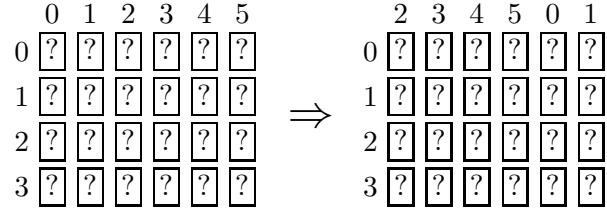

In a \textit{pile-shifting shuffle} on an $m \times n$ matrix, we shift the columns of the matrix by a random cyclic shift unknown to all parties, i.e. move each Column $\ell$ to Column $\ell+r$ for a uniformly random $r \in \mathbb{Z}/n\mathbb{Z}$ (where the indices are taken modulo $n$). See Fig. \ref{fig3}. This operation was introduced by Shinagawa et al. \cite{polygon}.

The pile-shifting shuffle can be implemented in real world by putting all cards in each column into an envelope, and then applying the random cut to the sequence of envelopes \cite{hindu}.

\section{Encoding Scheme of Shinagawa et al.}
\begin{figure}[H]
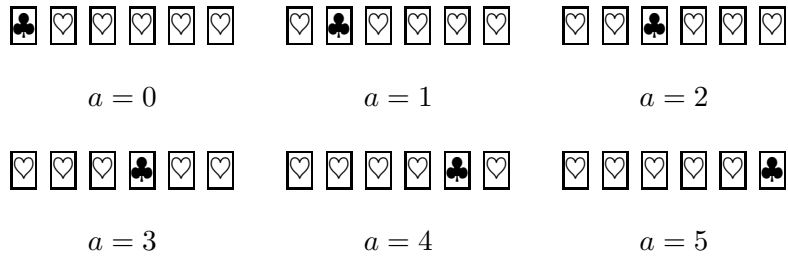

    \centering
    \begin{minipage}{3.5cm}
        \centering
        \mybox{$\clubsuit$} \mybox{$\heartsuit$} \mybox{$\heartsuit$} \mybox{$\heartsuit$} \mybox{$\heartsuit$} \mybox{$\heartsuit$} \\~\\
        $a=0$ \\~\\
				\mybox{$\heartsuit$} \mybox{$\heartsuit$} \mybox{$\heartsuit$} \mybox{$\clubsuit$} \mybox{$\heartsuit$} \mybox{$\heartsuit$} \\~\\
        $a=3$
    \end{minipage}
    \begin{minipage}{3.5cm}
        \centering
        \mybox{$\heartsuit$} \mybox{$\clubsuit$} \mybox{$\heartsuit$} \mybox{$\heartsuit$} \mybox{$\heartsuit$} \mybox{$\heartsuit$} \\~\\
        $a=1$ \\~\\
				\mybox{$\heartsuit$} \mybox{$\heartsuit$} \mybox{$\heartsuit$} \mybox{$\heartsuit$} \mybox{$\clubsuit$} \mybox{$\heartsuit$} \\~\\
        $a=4$
    \end{minipage}
    \begin{minipage}{3.5cm}
        \centering
        \mybox{$\heartsuit$} \mybox{$\heartsuit$} \mybox{$\clubsuit$} \mybox{$\heartsuit$} \mybox{$\heartsuit$} \mybox{$\heartsuit$} \\~\\
        $a=2$ \\~\\
				\mybox{$\heartsuit$} \mybox{$\heartsuit$} \mybox{$\heartsuit$} \mybox{$\heartsuit$} \mybox{$\heartsuit$} \mybox{$\clubsuit$} \\~\\
        $a=5$
    \end{minipage}
		\caption{Commitments of integers in $\mathbb{Z}/6\mathbb{Z}$ in the encoding scheme of Shinagawa et al.}
		\label{figA}
\end{figure}

Shinagawa et al. \cite{polygon} proposed an encoding scheme that uses a regular $n$-gon card to encode each integer in $\mathbb{Z}/n\mathbb{Z}$, which can be straightforwardly converted to the one using regular cards. In the converted scheme, an integer $a$ in $\mathbb{Z}/n\mathbb{Z}$ is encoded by a sequence $E_n(a)$ introduced in the previous section (see Fig. \ref{figA} for the case $n=6$).

We focus on three basic arithmetic protocols: the copy protocol, the addition protocol, and the multiplication protocol. These three protocols can be combined to compute any polynomial function $f: (\mathbb{Z}/n\mathbb{Z})^k \rightarrow \mathbb{Z}/n\mathbb{Z}$.

\subsection{Copy Protocol} \label{copy}
Given a sequence $A$ of $E_n(a)$, this protocol creates an additional copy of $A$ without revealing $a$. It was converted from a copy protocol of Shinagawa et al. \cite{polygon} which uses regular $n$-gon cards, and is also a generalization of a Boolean copy protocol of Mizuki and Sone \cite{mizuki09}.

\begin{enumerate}
	\item Reverse the $n-1$ rightmost cards of $A$, i.e. move each $(i+1)$-th leftmost card of $A$ to become the $i$-th rightmost card for $i=1,2,...,n-1$. This modified sequence, called $A'$, now encodes $-a$ (mod $n$).
	\item Construct a $3 \times n$ matrix $M$ by placing the sequence $A'$ in Row 0 and a sequence $E_n(0)$ in Row 1 and Row 2.
	\item Apply the pile-shifting shuffle to $M$. Note that Row 0 of $M$ now encodes $-a+r$ (mod $n$), and Row 1 and Row 2 now encode $r$ (mod $n$) for a uniformly random $r \in \mathbb{Z}/n\mathbb{Z}$.
	\item Turn over all cards in Row 0 of $M$. Locate the position of a \mybox{$\clubsuit$}. Suppose it is at Column $j$.
	\item Shift the columns of $M$ to the left by $j$ columns, i.e. move every Column $\ell$ to Column $\ell-j$ (where the indices are taken modulo $n$). Turn over all face-up cards.
	\item The sequences in Row 1 and Row 2 of $M$ now both encode $r-(-a+r) \equiv a$ (mod $n$), so we now have two copies of $A$ as desired.
\end{enumerate}

This protocol uses $n$ extra cards (one \mybox{$\clubsuit$} and $n-1$ \mybox{$\heartsuit$}s) in addition to the ones in $A$ and $A'$. Therefore, the total number of required cards is $3n$.

\subsection{Addition Protocol} \label{add}
Given sequences $A$ and $B$ of $E_n(a)$ and $E_n(b)$, respectively, this protocol computes the sum $a + b$ (mod $n$) without revealing $a$ or $b$. It was also converted from an addition protocol of Shinagawa et al. \cite{polygon} which uses regular $n$-gon cards.

\begin{enumerate}
	\item Reverse the $n-1$ rightmost cards of $A$, i.e. move each $(i+1)$-th leftmost card of $A$ to become the $i$-th rightmost card for $i=1,2,...,n-1$. This modified sequence, called $A'$, now encodes $-a$ (mod $n$).
	\item Construct a $2 \times n$ matrix $M$ by placing the sequence $A'$ in Row 0 and the sequence $B$ in Row 1.
	\item Apply the pile-shifting shuffle to $M$. Note that Row 0 and Row 1 of $M$ now encode $-a+r$ (mod $n$) and $b+r$ (mod $n$), respectively, for a uniformly random $r \in \mathbb{Z}/n\mathbb{Z}$.
	\item Turn over all cards in Row 0 of $M$. Locate the position of a \mybox{$\clubsuit$}. Suppose it is at Column $j$.
	\item Shift the columns of $M$ to the left by $j$ columns, i.e. move every Column $\ell$ to Column $\ell-j$ (where the indices are taken modulo $n$). Turn over all face-up cards.
	\item The sequence in Row 1 of $M$ now encodes $(b+r)-(-a+r) \equiv a+b$ (mod $n$) as desired.
\end{enumerate}

This protocol does not use any extra card other than the ones in $A$ and $B$. Therefore, the total number of required cards is $2n$.

\subsection{Multiplication Protocol} \label{multiply}
Given sequences $A$ and $B$ of $E_n(a)$ and $E_n(b)$, respectively, this protocol computes the product $a \cdot b$ (mod $n$) without revealing $a$ or $b$. It is a generalization of the protocol of Shinagawa and Mizuki \cite{triangle} to multiply two integers in $\mathbb{Z}/3\mathbb{Z}$, and is also a generalization of the Boolean AND protocol of Mizuki and Sone \cite{mizuki09}.

The intuition of this protocol is that we will create sequences $A_0,A_1,...,A_{n-1}$ encoding $0,a,2a,...,(n-1)a$ (mod $n$), respectively, and then select the sequence $A_b$ as an output.

\begin{figure}
\begin{center}
\begin{tikzpicture}
\node at (0.00,-1.10) {$A_0$};
\node at (0.50,-1.10) {$A_1$};
\node at (1.00,-1.10) {...};
\node at (1.80,-1.10) {$A_{n-1}$};

\draw[->] (0.00,-0.40) -- (0.00,-0.80);
\draw[->] (0.50,-0.40) -- (0.50,-0.80);
\draw[->] (1.50,-0.40) -- (1.50,-0.80);

\node at (0.00,0.00) {\mybox{?}};
\node at (0.50,0.00) {\mybox{?}};
\node at (1.00,0.00) {...};
\node at (1.50,0.00) {\mybox{?}};

\node at (0.00,0.70) {\vdots};
\node at (0.50,0.70) {\vdots};
\node at (1.00,0.70) {\vdots};
\node at (1.50,0.70) {\vdots};

\node at (0.00,1.20) {\mybox{?}};
\node at (0.50,1.20) {\mybox{?}};
\node at (1.00,1.20) {...};
\node at (1.50,1.20) {\mybox{?}};

\node at (0.00,1.80) {\mybox{?}};
\node at (0.50,1.80) {\mybox{?}};
\node at (1.00,1.80) {...};
\node at (1.50,1.80) {\mybox{?}};

\node at (0.00,2.70) {\mybox{?}};
\node at (0.50,2.70) {\mybox{?}};
\node at (1.00,2.70) {...};
\node at (1.50,2.70) {\mybox{?}};
\draw[->] (1.80,2.70) -- (2.2,2.7);
\node at (2.40,2.70) {$B$};

\draw[] (-0.30,-0.30) -- (-0.30,4.10);
\draw[] (-2.10,3.10) -- (2.00,3.10);

\node at (-0.60,0.00) {$n$};
\node at (-0.60,0.70) {\vdots};
\node at (-0.60,1.20) {2};
\node at (-0.60,1.80) {1};
\node at (-0.60,2.70) {0};
\node at (-1.60,1.35) {Row};

\node at (0.00,3.40) {0};
\node at (0.50,3.40) {1};
\node at (1.00,3.40) {...};
\node at (1.60,3.40) {$n-1$};
\node at (0.75,3.90) {Column};
\end{tikzpicture}
\caption{An $(n+1) \times n$ matrix $M$ constructed in Step 4}
\label{fig5}
\end{center}
\end{figure}
\begin{enumerate}
	\item Let $A_1=A$. If $n \geq 4$, we perform the following procedures for $n-3$ rounds. In each $i$-th round ($i=1,2,...,n-3$) when we already have sequences $A_1,A_2,...,A_i$, apply the copy protocol to create a copy of $A_1$ and a copy of $A_i$. Then, apply the addition protocol to the copy of $A_1$ and the copy of $A_i$. The resulting sequence, called $A_{i+1}$, encodes $a+ia \equiv (i+1)a$ (mod $n$).
	\item We now have sequences $A_1,A_2,...,A_{n-2}$. If $n \geq 3$, apply the copy protocol to create a copy of $A_1$ again. Reverse the $n-1$ rightmost cards of that copy, i.e. move each $(i+1)$-th leftmost card to become the $i$-th rightmost card for $i=1,2,...,n-1$. This modified sequence, called $A_{n-1}$, now encodes $-a \equiv (n-1)a$ (mod $n$).
	\item Arrange $n$ extra cards (which can be taken from the cards left from the copy protocol in Step 2 for $n \geq 3$) as a sequence $E_n(0)$, called $A_0$. We now have sequences $A_0,A_1,...,A_{n-1}$ as desired.
	\item Construct an $(n+1) \times n$ matrix $M$ by the following procedures (see Fig. \ref{fig5}).
	\begin{enumerate}
		\item In Row 0, place the sequence $B$.
		\item In each column $\ell=0,1,...,n-1$, place the sequence $A_\ell$ arranged vertically from Row 1 to Row $n$.
	\end{enumerate}
	\item Apply the pile-shifting shuffle to $M$.
	\item Turn over all cards in Row 0. Locate the position of a \mybox{$\clubsuit$}. Suppose it is at Column $j$.
	\item Select the sequence in Column $j$ arranged vertically from Row 1 to Row $n$. This is the sequence $A_b$ encoding $a \cdot b$ (mod $n$) as desired.
\end{enumerate}

In Step 1, in the $i$-th round we use $n$ extra cards in the copy protocol besides the cards in $A_1,A_2,...,A_i, B$, and the copies of $A_1$ and $A_i$, so the total number of cards is $(i+3)n+n \leq n^2+n$. In Step 2, we use $n$ extra cards in the copy protocol besides the cards in $A_1,A_2,...,A_{n-2}, B$, and the copy of $A_1$, so the total number of cards is $n^2+n$. Therefore, the total number of required cards for this protocol is $n^2+n$. Note that the special case $n=3$ works exactly like the multiplication protocol of Shinagawa and Mizuki \cite{triangle}, and the special case $n=2$ works exactly like the six-card AND protocol of Mizuki and Sone \cite{mizuki09}.

In summary, using the encoding scheme of Shinagawa et al. for $n=6$ requires 18, 12, and 42 cards for the copy, addition, and multiplication protocols, respectively.

\section{Encoding Scheme of Nishida et al.} \label{nishidascheme}
Nishida et al. \cite{nishida} developed a protocol to compute any $n$-variable Boolean function using $2n+6$ cards, where each bit $x$ in the inputs and output is encoded by $E_2(x)$. Their protocol also retains commitments of the inputs for further use. Hence, this protocol requires $2n$ cards for the inputs and two cards for the output, and actually uses four extra cards besides the ones in the inputs and output for the computation: two \mybox{$\clubsuit$}s and two \mybox{$\heartsuit$}s.

\begin{figure}[H]
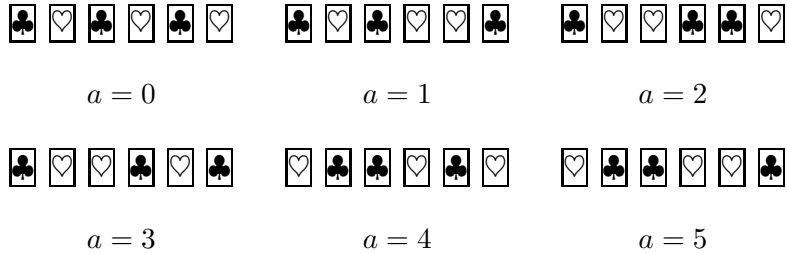

    \centering
    \begin{minipage}{3.5cm}
        \centering
        \mybox{$\clubsuit$} \mybox{$\heartsuit$} \mybox{$\clubsuit$} \mybox{$\heartsuit$} \mybox{$\clubsuit$} \mybox{$\heartsuit$} \\~\\
        $a=0$ \\~\\
				\mybox{$\clubsuit$} \mybox{$\heartsuit$} \mybox{$\heartsuit$} \mybox{$\clubsuit$} \mybox{$\heartsuit$} \mybox{$\clubsuit$} \\~\\
        $a=3$
    \end{minipage}
    \begin{minipage}{3.5cm}
        \centering
        \mybox{$\clubsuit$} \mybox{$\heartsuit$} \mybox{$\clubsuit$} \mybox{$\heartsuit$} \mybox{$\heartsuit$} \mybox{$\clubsuit$} \\~\\
        $a=1$ \\~\\
				\mybox{$\heartsuit$} \mybox{$\clubsuit$} \mybox{$\clubsuit$} \mybox{$\heartsuit$} \mybox{$\clubsuit$} \mybox{$\heartsuit$} \\~\\
        $a=4$
    \end{minipage}
    \begin{minipage}{3.5cm}
        \centering
        \mybox{$\clubsuit$} \mybox{$\heartsuit$} \mybox{$\heartsuit$} \mybox{$\clubsuit$} \mybox{$\clubsuit$} \mybox{$\heartsuit$} \\~\\
        $a=2$ \\~\\
				\mybox{$\heartsuit$} \mybox{$\clubsuit$} \mybox{$\clubsuit$} \mybox{$\heartsuit$} \mybox{$\heartsuit$} \mybox{$\clubsuit$} \\~\\
        $a=5$
    \end{minipage}
		\caption{Commitments of integers in $\mathbb{Z}/6\mathbb{Z}$ in the encoding scheme of Nishida et al.}
		\label{figB}
\end{figure}

We write each integer $a \in \mathbb{Z}/6\mathbb{Z}$ in its binary representation $a = (a_2,a_1,a_0)$, where $a=4a_2+2a_1+a_0$ and $a_0,a_1,a_2 \in \{0,1\}$. Each bit $a_i$ is encoded by $E_2(a_i)$, so we encode $a$ by a commitment of six cards consisting of $E_2(a_2)$, $E_2(a_1)$, and $E_2(a_0)$ arranged in this order from left to right (see Fig. \ref{figB}).

\subsection{Copy Protocol}
To copy a commitment of $a=(a_2,a_1,a_0)$, we apply the protocol in Section \ref{copy} to copy the sequences $E_2(a_0)$, $E_2(a_1)$, and $E_2(a_2)$ separately. Since the two extra cards used in that protocol can be reused in each computation, we use only two extra cards (one \mybox{$\clubsuit$} and one \mybox{$\heartsuit$}) besides the 12 cards encoding the inputs and outputs, resulting in the total of 14 cards.

\subsection{Addition Protocol}
Suppose we have integers $a=(a_2,a_1,a_0)$ and $b=(b_2,b_1,b_0)$.

Let $S = \{0,1\}^3 - \{(1,1,0),(1,1,1)\}$. Consider the following function $f_+: \{0,1\}^6 \rightarrow \{0,1\}^3$. Define
$$f_+(a_2,a_1,a_0,b_2,b_1,b_0) := (c_2,c_1,c_0),$$
where $(c_2,c_1,c_0)$ is the binary representation of $a+b$ (mod 6) if $(a_2,a_1,a_0),$ $(b_2,b_1,b_0) \in S$. We can define $f_+(a_2,a_1,a_0,b_2,b_1,b_0)$ to be any value if either $(a_2,a_1,a_0)$ or $(b_2,b_1,b_0)$ is not in $S$.

We apply the protocol of Nishida et al. \cite{nishida} to compute $c_0,c_1$, and $c_2$ separately. As explained at the beginning of Section \ref{nishidascheme}, this protocol retains the commitments of the inputs, and uses four extra cards (which can be reused in each computation) besides the ones in the inputs and outputs. Therefore, we use only four extra cards (two \mybox{$\clubsuit$}s and two \mybox{$\heartsuit$}s) besides the 18 cards encoding the inputs and outputs, resulting in the total of 22 cards.

\subsection{Multiplication Protocol}
Similarly to the addition protocol, consider a function $f_\times: \{0,1\}^6 \rightarrow \{0,1\}^3$ with
$$f_\times(a_2,a_1,a_0,b_2,b_1,b_0) := (c_2,c_1,c_0),$$
where $(c_2,c_1,c_0)$ is the binary representation of $a \cdot b$ (mod 6) if $(a_2,a_1,a_0),$ $(b_2,b_1,b_0) \in S$, and with $f_\times(a_2,a_1,a_0,b_2,b_1,b_0)$ being any value if either $(a_2,a_1,a_0)$ or $(b_2,b_1,b_0)$ is not in $S$.

Like in the addition protocol, we apply the protocol of Nishida et al. \cite{nishida} to compute $c_0,c_1$, and $c_2$ separately, which requires 22 cards in total.

In summary, using the encoding scheme of Nishida et al. requires 14, 22, and 22 cards for the copy, addition, and multiplication protocols, respectively.

\section{Our Encoding Scheme}
\begin{figure}[H]
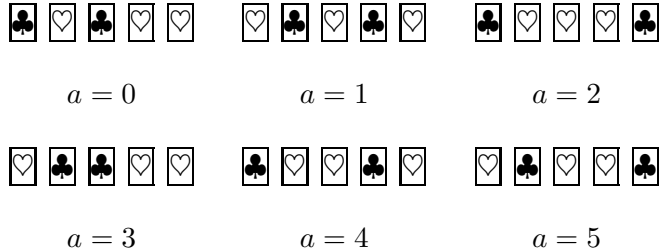

    \centering
    \begin{minipage}{2.93cm}
        \centering
        \mybox{$\clubsuit$} \mybox{$\heartsuit$} \mybox{$\clubsuit$} \mybox{$\heartsuit$} \mybox{$\heartsuit$} \\~\\
        $a=0$ \\~\\
				\mybox{$\heartsuit$} \mybox{$\clubsuit$} \mybox{$\clubsuit$} \mybox{$\heartsuit$} \mybox{$\heartsuit$} \\~\\
        $a=3$
    \end{minipage}
    \begin{minipage}{2.93cm}
        \centering
        \mybox{$\heartsuit$} \mybox{$\clubsuit$} \mybox{$\heartsuit$} \mybox{$\clubsuit$} \mybox{$\heartsuit$} \\~\\
        $a=1$ \\~\\
				\mybox{$\clubsuit$} \mybox{$\heartsuit$} \mybox{$\heartsuit$} \mybox{$\clubsuit$} \mybox{$\heartsuit$} \\~\\
        $a=4$
    \end{minipage}
    \begin{minipage}{2.93cm}
        \centering
        \mybox{$\clubsuit$} \mybox{$\heartsuit$} \mybox{$\heartsuit$} \mybox{$\heartsuit$} \mybox{$\clubsuit$} \\~\\
        $a=2$ \\~\\
				\mybox{$\heartsuit$} \mybox{$\clubsuit$} \mybox{$\heartsuit$} \mybox{$\heartsuit$} \mybox{$\clubsuit$} \\~\\
        $a=5$
    \end{minipage}
		\caption{Commitments of integers in $\mathbb{Z}/6\mathbb{Z}$ in our encoding scheme}
		\label{figC}
\end{figure}

In our encoding scheme, each integer $a \in \mathbb{Z}/6\mathbb{Z}$ is written as $(a_1,a_2)$, where $a_1$ and $a_2$ are remainders of $a$ when divided by 2 and 3, respectively. By Chinese remainder theorem, the value of $a$ is uniquely determined by $(a_1,a_2)$.\footnote{This can also be viewed as ring isomorphism $\mathbb{Z}/6\mathbb{Z} \cong (\mathbb{Z}/2\mathbb{Z}) \times (\mathbb{Z}/3\mathbb{Z})$.} We encode $a$ by a commitment of five cards, the first two cards being $E_2(a_1)$ and the last three cards being $E_3(a_2)$ (see Fig. \ref{figC}).

\subsection{Copy Protocol}
To copy a commitment of $a=(a_1,a_2)$, we apply the protocol in Section \ref{copy} to copy the sequences $E_2(a_1)$ and $E_3(a_2)$ separately. Since the extra cards used in that protocol can be reused in each computation, we use only three extra cards (one \mybox{$\clubsuit$} and two \mybox{$\heartsuit$}s) besides the ten cards encoding the inputs and outputs, resulting in the total of 13 cards.

\subsection{Addition Protocol}
Given $a=(a_1,a_2)$ and $b=(b_1,b_2)$, we have $a+b$ (mod 6$)= (c_1,c_2)$, where $c_1 = a_1+b_1$ (mod 2) and $c_2=a_2+b_2$ (mod 3). The values of $c_1$ and $c_2$ can be computed separately by applying the protocol in Section \ref{add}, which does not use any extra card. Therefore, the total number of required cards is ten.

\subsection{Multiplication Protocol}
Like the addition protocol, we have $a \cdot b$ (mod 6$)= (c_1,c_2)$, where $c_1 = a_1 \cdot b_1$ (mod 2) and $c_2=a_2 \cdot b_2$ (mod 3). The values of $c_1$ and $c_2$ can be computed separately by applying the protocol in Section \ref{multiply}, which in total uses six extra cards (two \mybox{$\clubsuit$}s and four \mybox{$\heartsuit$}s), so the total number of required cards is 16.

\subsubsection{Optimization} \label{opt}
By reusing cards, we can do a little better for the multiplication protocol. First, we compute $c_1$ using two extra cards (one \mybox{$\clubsuit$} and one \mybox{$\heartsuit$}). After the computation, we only use two cards (one \mybox{$\clubsuit$} and one \mybox{$\heartsuit$}) to encode $c_1$, so we now have four \textit{free} cards (two \mybox{$\clubsuit$}s and two \mybox{$\heartsuit$}s) that can be used in other computation. Since computing $c_2$ requires six extra cards (two \mybox{$\clubsuit$}s and four \mybox{$\heartsuit$}s), we actually need only two more \mybox{$\heartsuit$}s besides the four free cards we have. Therefore, in total we can use only four extra cards (one \mybox{$\clubsuit$} and three \mybox{$\heartsuit$}s), which reduces the number of required cards by two to 14.

In summary, using our encoding scheme for $n=6$ requires 13, 10, and 14 cards for the copy, addition, and multiplication protocols, respectively.

\section{Encoding Integers in Other Congruent Classes}
Our encoding scheme can be generalized to encode integers in $\mathbb{Z}/n\mathbb{Z}$ for any $n=p_1^{b_1}p_2^{b_2}...p_k^{b_k}$ such that $k>1$, where $p_1,p_2,...,p_k$ are different primes and $b_1,b_2,...,b_k$ are positive integers. For each $a \in \mathbb{Z}/n\mathbb{Z}$, let $a=(a_1,a_2,...,a_k)$, where each $a_i$ is the remainder of $a$ when divided by $p_i^{b_i}$. By Chinese remainder theorem, the value of $a$ is uniquely determined by $(a_1,a_2,...,a_k)$. We encode each $a_i$ by $E_{p_i^{b_i}}(a_i)$, so we use total of $\sum_{i=1}^k p_i^{b_i}$ cards for each commitment. We apply the protocols in Sections \ref{copy}, \ref{add}, and \ref{multiply} on each $E_{p_i^{b_i}}(a_i)$ separately to perform the copy, addition, and multiplication, respectively.

Let $m=\max_{i=1}^k p_i^{b_i}$, our encoding scheme requires $2 \sum_{i=1}^k p_i^{b_i}+m$ cards for the copy protocol, $2 \sum_{i=1}^k p_i^{b_i}$ cards for the addition protocol, and $2(\sum_{i=1}^k p_i^{b_i})+m^2-m$ cards for the multiplication protocol (before the optimization). By using the optimization technique in Section \ref{opt} (computing the smallest modulus first and reusing the free cards in larger modulii), we can slightly reduce the number of required cards for the multiplication protocol.

In comparison, the encoding scheme of Shinagawa et al. requires $3n$ cards for copy, $2n$ cards for addition, and $n^2+n$ cards for multiplication, while the encoding scheme of Nishida et al. requires $4\lceil \lg n \rceil+2$ cards for copy and $6\lceil \lg n \rceil+4$ cards for addition and multiplication.

The numbers of required cards for the copy, addition, and multiplication protocols for each applicable $n$ up to 20 are shown in Table \ref{table2}. Besides $\mathbb{Z}/6\mathbb{Z}$, our encoding scheme is also the currently best known schemes in $\mathbb{Z}/12\mathbb{Z}$ for the protocols of all three functions. For the addition protocol, our encoding scheme is the currently best known one for every such $n$.

\begin{table}
	\centering
	\begin{tabular}{|c|c|c|c|c|c|c|c|c|c|}
		\hline
		\multirow{2}{*}{\boldmath{$\mathbb{Z}/n\mathbb{Z}$}} & \multicolumn{3}{c|}{\textbf{Shinagawa et al. \cite{polygon}}} & \multicolumn{3}{c|}{\textbf{Nishida et al. \cite{nishida}}} & \multicolumn{3}{c|}{\textbf{Our scheme (\S 5)}} \\ \cline{2-10}
		& \textbf{Copy} & \textbf{Add.} & \textbf{Mult.} & \textbf{Copy} & \textbf{Add.} & \textbf{Mult.} & \textbf{Copy} & \textbf{Add.} & \textbf{Mult.} \\ \hline
		$\mathbb{Z}/6\mathbb{Z}$ & 18 & 12 & 42 & 14 & 22 & 22 & \textbf{13} & \textbf{10} & \textbf{14} \\ \hline
		$\mathbb{Z}/10\mathbb{Z}$ & 30 & 20 & 110 & \textbf{18} & 28 & \textbf{28} & 19 & \textbf{14} & 32 \\ \hline
		$\mathbb{Z}/12\mathbb{Z}$ & 36 & 24 & 156 & \textbf{18} & 28 & 28 & \textbf{18} & \textbf{14} & \textbf{23} \\ \hline
		$\mathbb{Z}/14\mathbb{Z}$ & 42 & 28 & 210 & \textbf{18} & 28 & \textbf{28} & 25 & \textbf{18} & 58 \\ \hline
		$\mathbb{Z}/15\mathbb{Z}$ & 45 & 30 & 240 & \textbf{18} & 28 & \textbf{28} & 21 & \textbf{16} & 33 \\ \hline
		$\mathbb{Z}/18\mathbb{Z}$ & 54 & 36 & 342 & \textbf{22} & 34 & \textbf{34} & 31 & \textbf{22} & 92 \\ \hline
		$\mathbb{Z}/20\mathbb{Z}$ & 60 & 40 & 420 & \textbf{22} & 34 & \textbf{34} & 23 & \textbf{18} & \textbf{34} \\ \hline
	\end{tabular}
	\medskip
	\caption{The number of required cards for copy, addition, and multiplication protocols (after the optimization) in each $\mathbb{Z}/n\mathbb{Z}$ using each encoding scheme, with the lowest number among each type of protocol boldfaced} \label{table2}
\end{table}

\section{Future Work}
We developed an encoding scheme for integers in $\mathbb{Z}/6\mathbb{Z}$ which allows us to perform the copy, addition, and multiplication using 13, 10, and 14 cards, respectively, which are the lowest numbers among the currently known protocols. We also generalized our encoding scheme to other rings of integers modulo $n$, including $\mathbb{Z}/12\mathbb{Z}$ where our protocols are the currently best known ones as well.

A challenging future work is to develop encoding schemes in $\mathbb{Z}/n\mathbb{Z}$ that requires fewer cards for other values of $n$, especially when $n$ is a prime, or prove the lower bound of the number of required cards for each $n$. For $\mathbb{Z}/6\mathbb{Z}$, we have to use at least four cards to encode each integer, no matter what the encoding scheme is (because three cards of two types can be rearranged in at most three ways). Hence, the trivial lower bound of the number of required cards for every protocol is eight.

Also, all results so far have been focused on using only two types of cards. An interesting question is that if we allow more than two types of cards, can we lower the number of required cards? (In particular, three different cards can be rearranged in six ways, so it might be possible to encode each integer in $\mathbb{Z}/6\mathbb{Z}$ with three cards.)

\end{document}